# Runtime QoS service for application-driven adaptation in network computing

Feras Al-Hawari
Electrical and Computer Engineering Department
Northeastern University
Boston, USA
feras_alhawari@yahoo.com

Elias Manolakos
Dept. of Informatics and Telecommunications
University of Athens
Athens, Greece
eliasm@di.uoa.gr

*Abstract*—A distributed application executing on a Network of Workstations (NOW) needs to be resource state aware to possibly adapt itself accordingly in order to keep satisfying the desired Quality of Service (QoS) demands throughout its lifespan. We implemented a QoS service to enable application-driven adaptation for performance and fault tolerance at runtime. The service is associated with lightweight middleware that monitors the state and load of all application entities (e.g., machines, tasks, and logical network links). Moreover, it makes its services available to an application task via an anonymous and simple to use QoS API. We present a Manager-Worker application that uses our fault tolerance QoS API to adapt for Worker faults in order to avoid application deadlock at runtime. Moreover, we show how a dynamic application-level scheduler can easily utilize the QoS API to find efficient schedules. Furthermore, we quantified the overhead of the QoS middleware in various scenarios to demonstrate that it has minor impact on the performance of the application it is servicing.

## I. INTRODUCTION AND RELATED WORK

The Networks of Workstations (NOW) resources are usually shared and heterogeneous, which makes the system state quite dynamic. So in order to deliver a desired QoS level (e.g., in terms of completion time or speedup ratio) to a network-computing (NC) application executing in such an environment, the characteristics of the resources before the application is launched (i.e., at startup time) must be considered to find an acceptable mapping of application tasks to machines. Furthermore, the application should remain resource state aware at runtime and possibly adapt itself accordingly in order to keep satisfying the targeted QoS demands throughout its lifecycle.

The task of monitoring the NOW resources can be delegated to an application-level QoS management system. The main requirements for such a system are: (1) finding a tasks-to-machines assignment that meets user-defined QoS levels at startup time and (2) making the dynamic system state easily accessible to a NC application to facilitate application adaptation at runtime. Our research group has developed such a flexible system for NC applications in the context of the ongoing JavaPorts (JP) framework project [1]. In this paper we present a QoS service and middleware designed to support application-driven adaptation for performance and fault tolerance at runtime. While, the details of the various startup-phase QoS management components are discussed in [2-4].

In order for an application to be able to adapt itself at runtime there must exist multiple execution paths for it to select from. For example, if similar Worker tasks of a Manager-Worker style application are replicated on multiple machines, the Manger task can adapt itself for fault tolerance by communicating with only responding Worker tasks. Furthermore, the Manager can adapt itself for performance by sending jobs or messages to the Workers running on the fastest machines or connected to it via the fastest network links, respectively.

The required mechanisms to support application adaptation at runtime are: (1) middleware to monitor and record the dynamic characteristics of the NOW resources and the application tasks, (2) an API to make the recorded information accessible to client applications, and (3) software modules to make adaptation decisions based on that information. The adaptation decisions can be made by the application itself (application-driven) or by system-level software modules that are transparent to the application (system-driven). Systems such as the Resource Monitoring System (Remos) [5], Globus Monitoring and Discovery System (MDS) [6-7], and Network Weather Service NWS [8], periodically monitor the dynamic characteristics of the underlying resources and provide an API to support application-driven adaptation at runtime i.e., the developer is responsible for implementing and adding the adaptation code in the application. They provide generic services to multiple applications and they are not aware of the configurations of their client applications (i.e., they mostly leave to the user the complex task of mapping the collected resource data to an application-level view). The API in these systems is not anonymous and requires the application's awareness of machine domain names as well as some network details.

On the other hand, system-driven adaptation frameworks, such as those in [9-21], support software modules that run along with the application in order to monitor its progress and to change its behavior when it does not meet the desired QoS levels. In such systems, the application code remains the same (i.e., no adaptation code is added) because the adaptation modules are transparent to the application. However, the developer is required to specify in a QoS profile when and how the system should change the application's behavior at runtime, which is not a trivial task and may prevent the user from modifying the adaptation criteria at run time.

We implemented a QoS service that enables application-driven adaptation at runtime. It implements an easy to use API that makes implementing various adaptation schemes an effortless task and eliminates the need to edit a QoS profile as in [9-21]. Unlike the NWS and Remos APIs, our API is anonymous, which makes the application code independent from the underlying resources i.e., the application can be configured to run on a new set of machines without any need to modify its code. The service is associated with middleware to monitor the dynamic state of the application tasks as well as the attributes of the machines and logical network links used by the application. The middleware is automatically configured, launched, and terminated along with the application it is servicing. It is lightweight because, unlike the NWS and Remos systems that provide services to multiple applications, it only provides services to its companion application (i.e., it does not waste cycles monitoring unused resources). Moreover, its awareness of the application configuration allows it to provide a task with services that enable it to easily access and adapt itself according to the attributes of any of the application entities.

The rest of the paper is organized as follows: the supported QoS service and API are defined in section II. The architecture of the various middleware modules is discussed in section III. The mechanisms to support fault tolerance are explained in section IV. The experiments used to validate our QoS API and middleware are discussed in section V. Finally, we summarize our findings in section VI.

## II. RUNTIME QOS SERVICE AND API

A JavaPorts (JP) application is composed of multiple, concurrent and possibly interacting tasks [1-3]. The structure of an application is abstracted as an Application Task Graph (ATG) (see Figure 1), where nodes correspond to tasks (components) and edges correspond to logical links between interacting tasks. Each task has its own predefined input-output ports. Two tasks may exchange messages via an edge (a point-to-point connection) using two peer ports (edge terminals). A JP task may communicate with a peer task using four synchronous/asynchronous read/write anonymous message passing operations [1]. In anonymous communications the name (and port) of the destination task does not need to be mentioned explicitly in the message passing method.

JP tasks get allocated to machines and several tasks may execute on the same machine (multi-tasking). The tasks-to-machines mapping can be easily modified at compile time to re-distribute the computational load without the need to re-code any part of the application logic (task location transparency). Each task is associated with either a Java or a Matlab software component and several tasks may share the same component implementation [2-3].

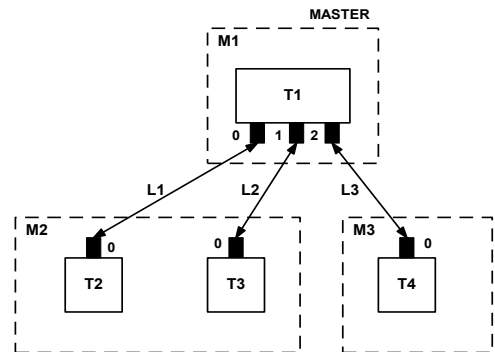

Figure 1. ATG for a Manager-Worker application in which the dashed rectangles represent logical machines, the solid rectangles represent tasks, and the solid lines represent the peer-to-peer logical links between the tasks.

The static/dynamic attributes of the ATG entities (i.e., machines, tasks, ports, and logical links) are needed by a running JP application to make adaptation decisions accordingly. Therefore, we implemented QoS middleware to measure and record the machines' attributes (see Table 1) as well as the logical links' throughput and latency. It also queries and stores the static information of all application entities (e.g., machine domain names, task variable names, port indices). Furthermore, it can detect machine and task states to provide services to support adaptation for fault tolerance.

TABLE 1: The static/dynamic machine attributes measured by the QoS middleware

| Machine Attribute | State | Comments |
|---|---|---|
| OSType | static | The OS type (e.g. Solaris) |
| CPUSpeed | static | The CPU clock rating (in MHz) |
| NumOfCPUs | static | The number of CPUs on a machine |
| Workload | dynamic | The average length of the run-queue of a machine, i.e. the average number of processes that are waiting in the ready-queue, plus the process(es) that is (are) currently executing on the machine's CPU(s); E.g., if the Workload of a single-CPU machine is 2, this means that there is two compute intensive processes sharing the same CPU i.e. one process running and one waiting in the ready-queue |
| EffectiveSpeed | dynamic | The effective CPU speed (in MHz) that a job will see when scheduled on a machine. It accounts for the contention effects of other jobs running on the machine. It is calculated analytically as follows: <br><br> EffectiveSpeed= factor*MinCPUSpeed |
| FreeRAMSize | dynamic | The free memory size in bytes |
| FreeSwapSize | dynamic | The free swap space size in bytes |
| *where:* | | |

$$factor = \begin{cases} 1, & \text{when } (1+Workload) \leq NumOfCPUs \\ \dfrac{NumOfCPUs}{(1+Workload)}, & \text{when } (1+Workload) > NumOfCPUs \end{cases}$$

MinCPUSpeed: The minimum clock rating (in MHz) of all CPUs on this machine.

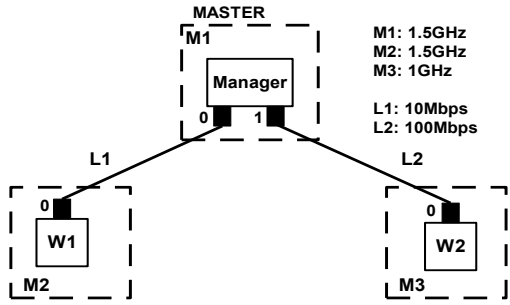

Figure 2: An ATG for a Manager-Worker application

The middleware stores the colected information in the ATG data structure in order to provide a calling task with hierarchical application-level views (objects). It supports four types of views: application (AppView), machine (MachView), task (TaskView), and port (PortView) views (see [4] for more information). An AppView object is a snapshot of all the application attributes and is represented as an array of MachView objects. Each MachView contains the attributes of its corresponding machine (e.g., state, workload, free memory size) and an array of TaskView objects representing the tasks on that machine. A TaskView contains task attributes (e.g., state, language, rank) and an array of PortView objects for each of the task ports. A PortView contains the attributes of all entities that a task sees through one of its ports i.e., the logical link, peer machine, peer task, and peer port.

The ability to get any of the application views enables a task to easily adapt according to the attributes of its neighboring or all application entities. For example, the Manager task in Figure 2 can simply access all the necessary information that it needs to adapt according to the attributes of its neighboring entities (e.g., peer Worker W1, peer machine M3, logical link L1) by obtaining its TaskView object. The Manager's TaskView contains the PortViews of ports 0 and 1. A PortView contains the attributes of its peer entities (e.g., the PortView of port 1 contains the attributes of logical link L2, machine M3, and Worker task W2). Based on that information, the Manager task can send a job or a message to the Worker on the peer machine with the highest CPUSpeed (i.e., machine M2) or over the logical link with highest throughput (i.e., link L2), respectively.

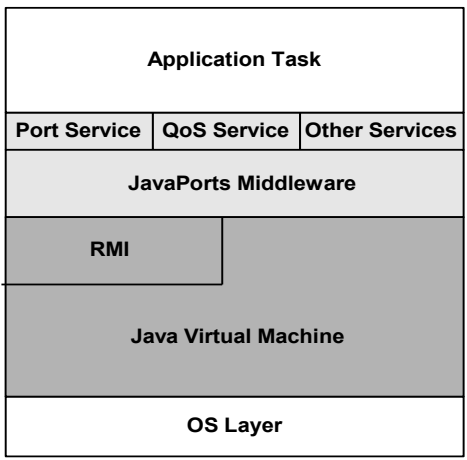

Figure 3: The JavaPorts middleware layer stack

TABLE 2: QoS API methods examples (see [4] for the full API specification)

| method signature | method invocation example |
|---|---|
| public AppView GetAppView(); | AppView av = qosservice.GetAppView(); |
| public int GetMeasStamp(); | int stamp= qosservice.GetMeasStamp(); |
| public int GetTaskRank(); | int r = qosservice.GetTaskRank(); |
| public void StopMonitoring(); | qosservice.StopMonitoring(); |

The JP framework makes the collected information available to an application task via an intermediary QoSService object (shown in Figure 3), which implements all the QoS API methods [4] and serves as an interface between an application task and the underlying QoS middleware. Thus, a task may for example obtain its AppView by invoking the corresponding QoS API method on its QoSService object. The QoS API method retrieves the AppView object by transparently interacting with the underlying QoS middleware. The supported QoS API methods (see [4] for the complete API specification) are categorized as follows:

- **View retrieval methods** to allow a calling task to get handles to a view object (e.g., the GetAppView method in Table 2).
- **Utility methods** to allow a task to find the index of the port to a best attribute resource, sort its PortView array in descending/ascending order by a machine or link attribute, or get the latest measurements stamp (e.g., the GetMeasStamp method in Table 2).
- **General methods** to allow a calling task to get general static information such as its rank and variable name, the rank of its machine, and the number of tasks/machines in the application (e.g., the GetTaskRank method in Table 2).
- **Middleware management methods** to enable a task to stop/resume the QoS monitoring modules (e.g., the StopMonitoring method in Table 2) and modify the middleware preferences at runtime (e.g., resource monitoring frequency).

III. MIDDLEWARE ARCHITECTURE

The QoS middleware is tightly coupled, and automatically launched and terminated along, with the application it is servicing. Therefore, the configuration of, and interactions between, its various modules are based on the ATG of the companion application. So based on the application configuration in Figure 1, the middleware architecture and interactions between its various modules and objects are as shown in Figures 4 and 5.

There are two modes that the middleware operates on and that coexist (concurrent). In the first mode (i.e., monitoring mode) the QoS managers measure and record (in the QoS data objects) the attributes of the machines, tasks, and logical links (see Figure 4). While in the second mode (i.e., servicing mode) a task may query the middleware for a service by invoking a QoS API method on its QoSService object (see Figure 5). The QoS API method retrieves the required attribute values from the corresponding QoS data object(s) to deliver the desired service to the task.

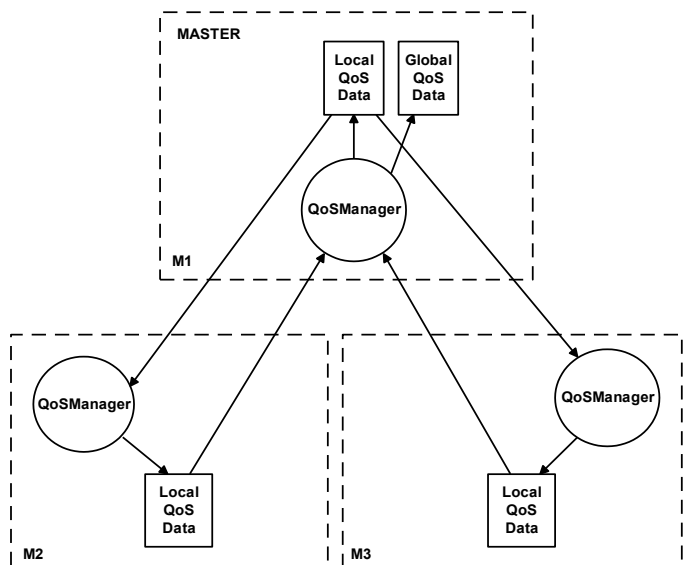

Figure 4: Interactions between the QoSManagers and the data objects to collect/store the QoS related data. Circular nodes represent threads, solid rectangles represent objects, and dashed rectangles represent machine boundaries. The arrow directions indicate the type of read/write interaction between threads and objects.

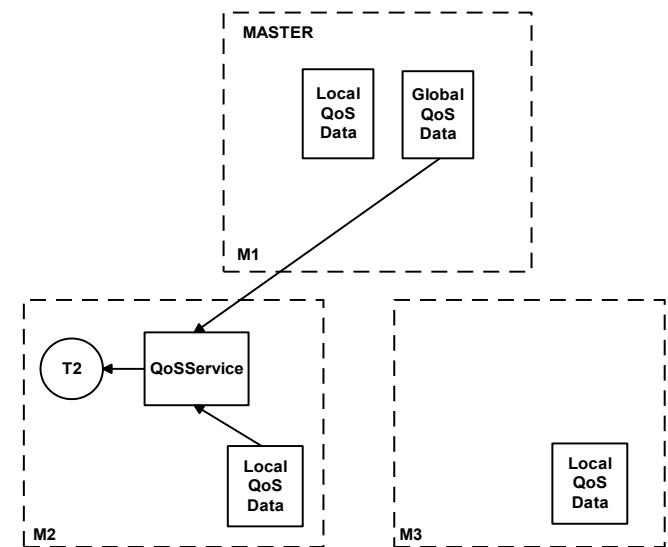

Figure 5: Interactions between user task T2 and its QoSService object, and between the QoSService object and the shared QoS data objects during a service request.

The lightweight QoSManager threads are the basic middleware modules and are associated with each machine in the ATG of the companion application (i.e., the number of QoSManager threads is equal to the number of machines in the ATG). A QoSManager is responsible for: (1) instantiating a LocalQoSData object on its machine, (2) retrieving the static information of its machine (e.g. OS type), (3) measuring the dynamic attributes of its machine (e.g., workload), (4) retrieving the attributes of its peer machines (if any), (5) measuring the throughput/latency of its designated logical links (if any) according to a token passing algorithm discussed in [4], and (6) storing the collected data in its MachView object, which is encapsulated in the LocalQoSData object. Moreover, the QoSManager on the MASTER machine constructs and updates the AppView object, which is stored inside the GlobalQoSData object on its machine.

JP designates one task on each machine (based on the location of the involved tasks in the ATG) to automatically launch the QoSManager on that machine. The Java Remote Method Invocation (RMI) technology is utilized to make the Local/Global QoSData objects accessible to local and remote QoSManagers. A QoSManager makes a QoSData object shared by registering it in the rmiregistery on its machine. An rmiregistry is a simple naming facility that allows remote/local clients to get a reference to a shared object so that they can invoke its methods.

Based on Figure 1 and Figure 4, the QoSManager on the MASTER machine measures the attributes of machine M1 and retrieves the attributes of machines M2 and M3 in order to construct its MachView and AppView, respectively. While, the QoSManager on machine M2 measures the attributes of machine M2 and retrieves the attributes of machine M1 (since tasks T2 and T3 are both connected to task T1 on machine M1) in order to construct its MachView. Finally, the QoSManager on machine M3 constructs its MachView by measuring the attributes of its machine M3 and retrieving the attributes of machine M1 (since task T4 is connected to task T1 on machine M1). The QoSManagers on machines M2 and M3 do not need to access the LocalQoSData on machines M3 and M2 respectively because tasks T2 and T3 have no peer tasks allocated to machine M3 and task T4 has no peer tasks allocated to machine M2.

The QoSManagers and QoS data objects are all hidden from the companion application. As we mentioned earlier, an intermediary QoSService object implements a QoS API to enable a task to access the desired QoS data by invoking various API methods. Each task instantiates a QoSService object at startup (JP automatically adds the QoSService object instantiation code in all the task templates [4]) in a similar manner as the other JP message passing Port objects [1]. Therefore, the QoSService object gives the user the same feel as the other available Port objects, with the exception that the message passing API methods cannot be invoked on it (separation of concerns). A QoSService object has access to the application's configuration information as well as the needed local/global QoS data objects so it can deliver the requested services to its task. Moreover, the Java programming language monitor mechanism is used to synchronize the concurrent writes or reads of multiple threads to or from the same attribute (e.g., two QoSService methods, or a QoSService method and a QoSManager contending for the same attribute in a GlobalQoSData or a LocalQoSData object respectively).

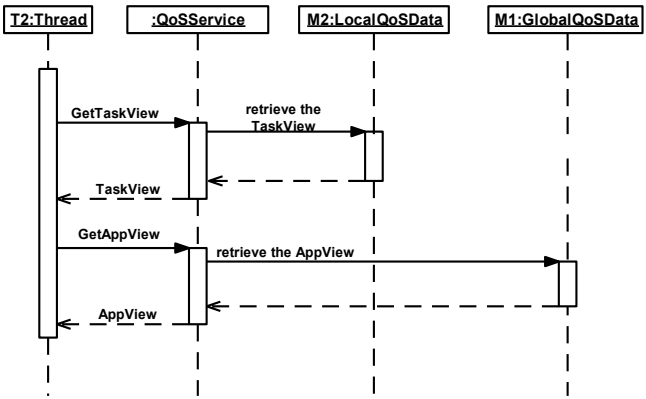

Figure 6: The various interactions between user task T2 and its QoSService object, and between the QoSService object and the shared QoS data objects when the GetTaskView() and GetAppView() methods are invoked.

The interactions between task T2 and the QoSService object as well as the QoSService object and the shared QoS data objects are shown in Figure 5. The QoSService object of task T2 has access to the LocalQoSData object (on its machine M2) and the GlobalQoSData object (on the MASTER machine M1). The fact that it is interacting with several objects in Figure 5 does not mean that they are all used at the same time. The read interactions with the LocalQoSData objects and the GlobalQoSData are used to retrieve the machine and application views respectively. The QoSService object of task T2 does not need to access any remote LocalQoSData objects because its QoSManager retrieves and stores the remote data in the LocalQoSData object.

Moreover, the interaction diagram in Figure 6 shows the various interactions between user task T2 and its QoSService object, and between the QoSService object and the shared QoS data objects when the GetTaskView() and GetAppView() methods [4] are invoked. For example, the GetTaskView() API method makes the handle of the TaskView of task T2 available to task T2 after retrieving it from the MachView object that is encapsulated in the LocalQoSData object on machine M2. Moreover, the GetAppView() API method invokes a method on the GlobalQoSData handle stored in its QoSService object in order to retrieve the AppView object and then pass its handle to the calling task.

## IV. FAULT TOLERANCE SUPPORT

The middleware can detect machine and task faults at runtime. Moreover, it provides various QoS API methods [4] to enable a task to check the state of its peer machines/tasks, before communicating with them, in order to avoid deadlock.

Machine faults (e.g., crashes) are detected using the Unix/Linux ping utility. If a machine is responding to the ping utility it is considered in the up state, otherwise it is considered in the down state. We also repeat each ping twice to avoid false alarms.

```
   // Get peer-port task state
1. state = portView.GetPeerTaskAttributeValue( ResourceAttribute.TASKSTATE ) ;

   // Get peer-port machine state
2. state = portView.GetPeerMachAttributeValue( ResourceAttribute.MACHSTATE , false ) ;

   // Get MachView machine state
3. state = machView.GetAttributeValue( ResourceAttribute.MACHSTATE , false ) ;

   // Get TaskView task state
4. state = taskView.GetAttributeValue( ResourceAttribute.TASKSTATE ) ;
```
Figure 7: QoS API methods to get a machine or task state.

On the other hand, a task's state is detected as follows. Initially, all tasks in the ATG are considered in the init state. When a task is up (i.e., launched), the middleware changes the task's state in the local/global QoS data objects to running. In addition, the QoSManager gets and records the process ID (PID) of the task. Before a task exits, the middleware sets its state in its corresponding local/global QoS data objects to completed (i.e., exited normally). When a running task exits abnormally (e.g., killed or crashed) it is considered to be in the dead state. The middleware detects the dead state by getting the task's current state and PID from the Local or GlobalQoSData objects. Then, if the task is in the running state it pings its machine to check its state. If the machine is down the task is considered dead, else the Unix/Linux ps utility is used to get the PIDs of the tasks running on that machine. The task is considered running, if its PID is in the PIDs list, else it is considered dead.

There are several methods in the QoS API that allow an application task to get the state of a machine or a task. For example, given a PortView object, a task may get the state of the peer-port task or machine as shown at lines 1 and 2 in Figure 7, respectively. Moreover, given a MachView or a TaskView object, a task may get the machine or task state as shown at lines 3 and 4 in Figure 7, respectively.

The pseudo code in Figure 8 demonstrates how the Manager task in Figure 2 may easily use the fault tolerance QoS API methods to check the state of a Worker task, before communicating with it, in order to avoid deadlock. At lines 2 and 3, the Manager task obtains the state of a Worker task. At line 4, it checks if the ready Worker is in the dead state. If not, it can communicate with the Worker (at line 5), otherwise it marks that Worker dead and it will not communicate with it again (at line 6).

```
1.  for (each worker p) {
2.      PortView  portView = qosservice_.GetPortView( p ) ;
3.      workerState = portView.GetPeerTaskAttributeValue (
                                     ResourceAttribute.TASKSTATE ) ;

4.      if ( workerState != -1 ) { // i.e. worker is running
5.          // communicate with worker
        }
        else{
6.          // mark this worker as dead and don't communicate with it
        }
    }
```
Figure 8: Pseudo code that demonstrates how the Manager task (in Figure 2) may use the fault tolerance QoS API methods to check the Workers states.

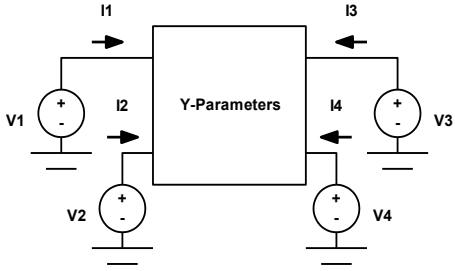

Figure 9: 4-port circuit model example

## V. VALIDATION AND RESULTS

We implemented a Manager-Worker application to calculate the time domain (TD) currents entering an N-port circuit (see Figure 9) characterized by a given matrix of Y-Parameters (admittances). Assuming the frequency domain (FD) voltage stimuli at the ports are known, the FD currents entering each port can be calculated according to equation 1. Then the Inverse Fast Fourier Transform (IFFT) can be used to transform each FD current to TD.

$$\begin{aligned} I_1(j\omega) &= Y_{11}V_1(j\omega) + Y_{12}V_2(j\omega) + \ldots + Y_{1N}V_N(j\omega) \\ I_2(j\omega) &= Y_{21}V_1(j\omega) + Y_{22}V_2(j\omega) + \ldots + Y_{2N}V_N(j\omega) \\ &\vdots \\ I_N(j\omega) &= Y_{N1}V_1(j\omega) + Y_{N2}V_2(j\omega) + \ldots + Y_{NN}V_N(j\omega) \end{aligned} \quad (1)$$

where:
N: number of ports.
$I_n(j\omega)$: frequency domain current at port n, $1 \leq n \leq N$.
$V_n(j\omega)$: frequency domain voltage at port n, $1 \leq n \leq N$.
$Y_{ij}$: the ij-th element in the admittance matrix.

### A. Adaptive manager-worker application

We studied non-adaptive and adaptive application instances in this experiment. The Assign (Manager-Worker) programming paradigm [23] is used in both instances. Based on this paradigm, the Manager distributes the data to multiple workers and determines how much work each of the workers performs. The partitioned data is sent to the workers that execute the same code. A Worker performs its work and sends the result back to the Manager. We assume that there is only one Worker on every machine except the MASTER machine (see Figure 2).

In the non-adaptive instance the Manager partitions the work evenly among the workers. Hence, if N is the number of ports and W is the number of workers, then each Worker calculates L=N/W port currents (according to Equation 1). The Manager starts by distributing $V_n(j\omega)$ voltage vectors and the corresponding Y coefficients to all workers based on a scatter communication pattern. Each Worker calculates its designated L FD currents and then their IFFT. The Manager collects the resulting L TD current vectors sent by each Worker.

In the adaptive instance, the Manager partitions the load among workers according to the workload of their machines. Based on that, the least loaded workers compute more currents than the heavily loaded ones. Assuming the workers are running on a set of homogeneous machines, we can formulate the partitioning problem similarly to AppLeS [22-23] as follows:

$$EW_m = L_m * Workload_m \quad (2)$$

where:
- $EW_m$: expected Workload of machine m after assigning $L_m$ currents to it.
- $L_m$: the number of currents assigned to machine m
- $Workload_m$: The measured Workload of machine m.

To balance the load among W workers, we solve the system of linear equations 3 and 4 for $L_m$.

$$EW_1 = EW_2 = \ldots = EW_M \quad (3)$$
$$L_1 + L_2 + \ldots + L_M = W \quad (4)$$

The complexity of solving the system of equations is $O(M^3)$, where M is the number of used machines (i.e., workers). Usually, the number of used machines is small, which makes solving these equations scalable and fast. The pseudo code for the Manager and Worker tasks of this application are shown in Figure 10.

In this experiment, we used six workers launched on a set of 333MHz machines running UNIX and connected by 100Mbps network links. In each run we varied the vector size as follows: 500, 1000, 1500, and 2000. We varied N as follows: 12, 30, and 60. We repeated each run three times, under the same load conditions, and reported the average completion time.

In order to conduct a fair evaluation of the adaptive and non-adaptive application instances we ran them under controlled load conditions. So, before conducting an experiment we select a set of lightly loaded machines and then we selectively overload some of the machines by executing a load generation application on them. The load generation application simply launches a job or several jobs that compute the FFT of a large size array of doubles (e.g., 5k) on the selected machine(s). When a job completes execution, the load generation application re-launches a new job on the corresponding machine after some specified delay.

We compared the application's completion time of the adaptive and non-adaptive instances under two load conditions: (1) Load1: Three machines are lightly loaded and the other three machines with a static workload of two introduced by the load generator, and (2) Load2: Two machines are lightly loaded, two machines with static workload of two introduced, and two machines with variable load introduced. The variable load alternates between a maximum workload of two to a zero workload every 6 seconds.

```
public synchronized void run ()
{
  - set N     // number of ports i.e. currents to calculate
  - set W     // number of workers
  - set vectorSize // the length of each current array

  L = calcL ( N, W ) ;

  - send Y, L[ m ], vectorSize, and all Vn(jw) vectors to each worker m

  n = 0;
  for ( m = 0 ; m < W ; m++ ) {
    for ( k = 0 ; k < L[ m ] ; k++ ) {
      I[ n ] = (double [ ] ) port_[ m ].SyncRead( k ) ;
      n++ ;
    }
  }
}
```

(a)

```
public synchronized void run ()
{
  - get Y, L[ m ], vectorSize, and all Vn(jw) vectors from manager
  - calculate L[ m ] time domain currents
  - send the calculated currents back to the manager
}
```

(b)

```
public int [ ] calcL ( int N, int W ) {
  if ( QoS support ) {
    newStamp = qosservice_.GetMeasStamp( );

    if ( newStamp != oldStamp ) {
      oldStamp = newStamp ;
      taskView = qosservice_.GetTaskView( ) ;
      portViews = taskView.GetPortViews( ) ;

      for ( m = 0 ; m < W ; m++ ) {
        Workload[ m ] = portViews[ m ].GetPeerMachAttributeValue(
                           ResourceAttribute.WORKLOAD, false ) ;
      } // for m
    }
    - Formulate the system of linear equations
    - Solve the system of linear equation using LU decomposition to find L
  }
  else { // no QoS support
    for ( m = 0 ; m < W ; m++ ) {
      L[ m ] = N / W ;
    }
  }
  return L ;
}
```

(c)

Figure 10: Pseudo code for the: (a) Manager task, (b) Worker task, and (c) calcL() method.

TABLE 3: Results under Load1: (a) average elapsed times in minutes, and (b) difference between the non-adaptive and adaptive results; where Diff equals the difference between the no and with adaptation elapsed times i.e. Diff = (No –With).

| | Average Elapsed Time in minutes | | | | | | | |
|---|---|---|---|---|---|---|---|---|
| | Vector size 500 | | Vector size 1000 | | Vector size 1500 | | Vector size 2000 | |
| N | No QoS | With QoS | No QoS | With QoS | No QoS | With QoS | No QoS | With QoS |
| 12 | .193 | .173 | .72 | .546 | 1.62 | 1.225 | 2.88 | 2.08 |
| 30 | .65 | .36 | 2.51 | 1.35 | 5.51 | 3.15 | 9.6 | 5.35 |
| 60 | 1.48 | .88 | 5.25 | 3.16 | 12.1 | 6.58 | 20.8 | 10.9 |

(a)

| | Vector size 500 | Vector size 1000 | Vector size 1500 | Vector size 2000 |
|---|---|---|---|---|
| N | Diff/No % | Diff/No % | Diff/No % | Diff/No % |
| 12 | 10.3% | 26% | 25% | 28% |
| 30 | 45% | 46% | 43% | 44% |
| 60 | 40% | 39.8% | 46% | 47.5% |

(b)

TABLE 4: Results under Load2: (a) average elapsed times in minutes, and (b) difference between the non adaptive and adaptive results; where Diff = (No – With).

| | Average Elapsed Time in minutes | | | | | | | |
|---|---|---|---|---|---|---|---|---|
| | Vector size 500 | | Vector size 1000 | | Vector size 1500 | | Vector size 2000 | |
| N | No QoS | With QoS | No QoS | With QoS | No QoS | With QoS | No QoS | With QoS |
| 12 | 0.195 | 0.17 | .83 | .53 | 1.63 | 1.18 | 3 | 2.04 |
| 30 | 0.61 | 0.38 | 2.48 | 1.45 | 5.25 | 3.33 | 9.4 | 5.5 |
| 60 | 1.54 | .88 | 5.6 | 3.6 | 12.46 | 7.38 | 21.9 | 12.6 |

(a)

| | Vector size 500 | Vector size 1000 | Vector size 1500 | Vector size 2000 |
|---|---|---|---|---|
| N | Diff/No % | Diff/No % | Diff/No % | Diff/No % |
| 12 | 13% | 36% | 27.5% | 32% |
| 30 | 38% | 41.5% | 37% | 41.4% |
| 60 | 43% | 36% | 41% | 42.5% |

(b)

The results of this experiment under Load1 and Load2 are shown in Tables 3 and 4, respectively. In both scenarios, the adaptive instance always outperformed the non-adaptive instance because it assigned more/less work to the lightly/heavily loaded machines rather than splitting the work among all machines regardless of their Workload. Moreover, the margin of improvement was more for N=30 and N=60 than when N=12, since when N is larger the non-adaptive instance assigned more work to the heavily loaded machines, which degraded its performance relatively to the adaptive instance.

### B. Measuring the middleware overhead

In this experiment, we used the non-adaptive instance of the Manager-Worker application in order to assess how disruptive and performance stealing monitoring is. We compared the application's completion time when the QoS monitoring is turned off to its time when the QoS monitoring is turned on but the QoS API is not used (i.e. just the message passing API is used). When QoS is on, the QoSManagers run concurrently with application, but the application will not use any of the QoS services i.e., the Manager will split the work evenly among the workers regardless of state of the machines. When QoS is off, the Manager will do the same thing, but in this case the QoS monitoring is turned off and the QoSManagers will not run concurrently with the application.

TABLE 5: The QoS system overhead as seen from the application perspective when N = 60, and W = 6: (a) average elapsed times in minutes, and (b) the difference between the results when the QoS support is off/on; where Diff = (No –With).

| | Average Elapsed Time in minutes | | | | | | | |
|---|---|---|---|---|---|---|---|---|
| | Vector size 500 | | Vector size 1000 | | Vector size 1500 | | Vector size 2000 | |
| period | No QoS | With QoS | No QoS | With QoS | No QoS | With QoS | No QoS | With QoS |
| 5 sec | .765 | .858 | 2.71 | 2.91 | 5.99 | 6.41 | 10.4 | 10.86 |
| 30 sec | .765 | .778 | 2.71 | 2.82 | 5.99 | 6.08 | 10.4 | 10.73 |
| 60 sec | .765 | .773 | 2.71 | 2.765 | 5.99 | 6.05 | 10.4 | 10.58 |

(a)

| | Vector size 500 | Vector size 1000 | Vector size 1500 | Vector size 2000 |
|---|---|---|---|---|
| period | Diff/No % | Diff/No % | Diff/No % | Diff/No % |
| 5 sec | -12% | -7.3% | -7% | -4.4% |
| 30 sec | -1.7% | -4% | -1.5% | -3.2% |
| 60 sec | -1% | -2% | -1% | -1.7% |

(b)

We used six workers executing on lightly loaded 333MHz machines. We set N to 60 and varied the vector sizes as follows: 500, 1000, 1500, and 2000. We compared the application's completion time when the QoS monitoring is off to its time when the QoSManagers are monitoring the resources every 5, 30, and 60 seconds. Moreover, we ran each case three times and reported their average times.

The results in Table 5 show that the QoS monitoring impact on the application performance decreases as the monitoring period increases (i.e., the middle is less intrusive). Moreover, when the monitoring period is greater than or equal 30 seconds the system overhead is very small and did not exceed 4% of the application's overall time in all cases. However, when the QoS monitoring is more intrusive (i.e. when the monitoring period is 5 seconds) the impact on the application's time reached 12% for the smaller vector sizes and was about 6.5% on average for the larger message sizes. When the vector size is small the application time is very short (less than a minute), so if the monitors use a few seconds during that time frame then we see a higher impact on the application's performance. Thus, the benefit of using the QoS service far exceeds its impact on the application's performance even when the QoS monitoring is very intrusive. Furthermore, when the middleware is used in a non-intrusive manner (e.g. monitoring periods >= 30 seconds) it will have a very small impact on the application's performance even if all the resources are homogeneous.

## VI. SUMMARY

We introduced a QoS middleware that runs along with the companion application to monitor its resources and to support adaptation for performance and application fault tolerance. It provides an application task with a QoS service that implements an easy to use and anonymous QoS specific API. The service enables a task to assess, and adapt itself according to, the dynamic load/state conditions of the neighboring or of all application entities. We showed how the QoS API could be easily used to implement various scenarios to adapt for performance or fault tolerance, which also lead to more efficient and robust applications. We also demonstrated that the middleware is lightweight and has minor impact on the performance of the companion application.